\documentclass[11pt,a4paper]{article}
\usepackage[margin=2.5cm]{geometry}
\usepackage[T1]{fontenc}
\usepackage{lmodern}
\usepackage{microtype}
\usepackage{graphicx}
\usepackage[hyphens]{url}
\usepackage[numbers,sort&compress]{natbib}
\usepackage{tikz}
\usetikzlibrary{arrows.meta}
\usepackage[colorlinks=true,allcolors=blue!50!black]{hyperref}

\title{Engineering process plants as code in the age of agentic AI}
\author{
  Artur M. Schweidtmann\thanks{Corresponding author. E-mail: \href{mailto:a.schweidtmann@tudelft.nl}{a.schweidtmann@tudelft.nl}}\\
  {\small Process Intelligence Research Group, Department of Chemical Engineering,}\\
  {\small Delft University of Technology, Delft, The Netherlands}\\[2ex]
  Davy Demeyer\\
  {\small Acceleer, Zaventem, Belgium}
}
\date{Preprint, October 2026}

\begin{document}
\maketitle

\begin{abstract}
Agentic artificial intelligence (AI) could one day shorten process plant engineering from years to weeks. Today, however, plant information is scattered across documents and tools and must be carried between them by hand, so an AI agent placed into current workflows inherits the same silos. In this Comment, we argue for Plant-as-Code: defining the engineering of a plant as a version-controlled collection of machine-readable files that follow, where possible, existing standards such as DEXPI and IFC. On this foundation, Design-Ops extends DevOps principles to plant engineering, linking every engineering item to the information it derives from and replacing manual checking with automated tests. Agents can then coordinate routine work across disciplines while deterministic tools perform the calculations. Verification sets the pace: agents propose, tools check, and engineers review and authorize every change. We outline the risks of this shift and the research, industry and policy actions it requires.
\end{abstract}

New medicines, battery materials and low-carbon chemicals all pass
through the same narrow gate before they reach the market. Someone
has to design, build and commission a plant that makes them, and the
engineering alone routinely takes years and a substantial share of
the investment. Much of this effort goes into moving information
between documents, checking it and moving it again. At the same time,
a generation of experienced plant engineers is retiring in many
industrialized countries and taking tacit knowledge with it. These
pressures are felt most strongly where engineering hours are
expensive, but the opportunity is broader. With capable AI support,
regions with limited engineering expertise could design plants that
today depend on scarce specialists.

Agentic artificial intelligence~(AI), in which language models plan
tasks and call software tools, could change this. Early studies
already show AI pipelines that turn process descriptions into
flowsheets and simulations at a level comparable to expert
engineers~\cite{Laub2026TextToFlowsheet}, and agents that coordinate
several engineering subtasks~\cite{Rupprecht2026MultiAgent}. In our
view, model capability is only one obstacle to wider impact. Equally
important is the way plant engineering data are stored and exchanged
today. In this Comment, we argue that plants should be defined as
code, that the engineering lifecycle should be organized as
Design-Ops, and that verification will set the pace of progress.

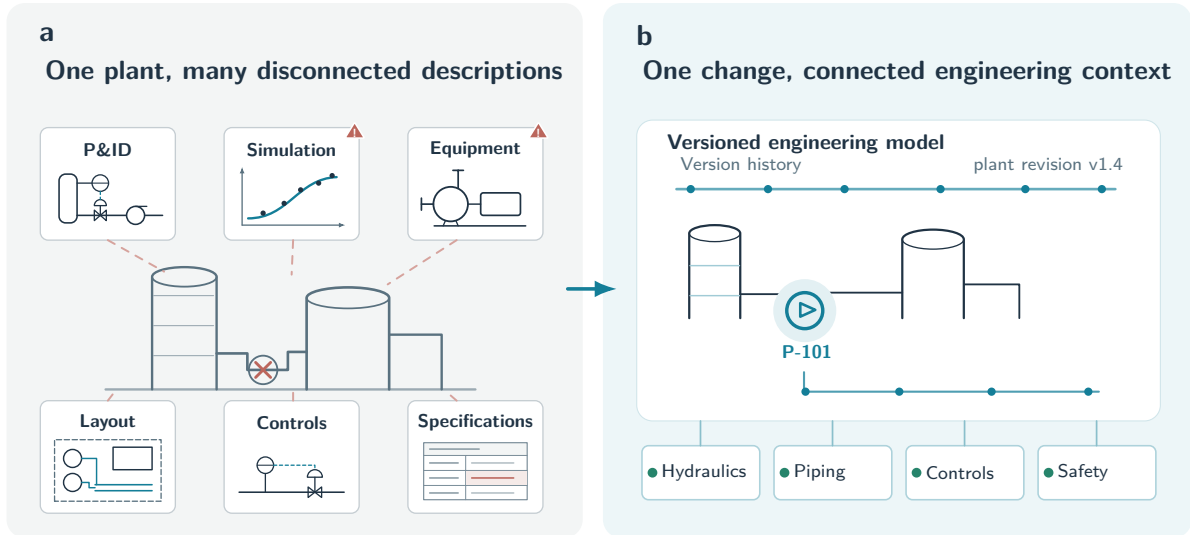
\begin{figure}[t]
  \centering
  \resizebox{\textwidth}{!}{
\begingroup\renewcommand{\familydefault}{\sfdefault}\normalfont
\begin{tikzpicture}[x=1cm,y=1cm, line cap=round, line join=round, font=\sffamily]
\definecolor{figink}{HTML}{203344}
\definecolor{figmuted}{HTML}{59717F}
\definecolor{figblue}{HTML}{147E98}
\definecolor{figpale}{HTML}{EDF6F8}
\definecolor{figgray}{HTML}{F3F5F5}
\definecolor{figred}{HTML}{BB665C}
\definecolor{figgreen}{HTML}{26846B}
\path[use as bounding box] (0,.95) rectangle (16,8.4);
\fill[figgray,rounded corners=7pt] (0.1,1.07) rectangle (7.78,8.2);
\fill[figpale,rounded corners=7pt] (8.05,1.07) rectangle (15.9,8.2);
\node[anchor=west,font=\bfseries\normalsize,text=figink] at (.40,7.78) {a};
\node[anchor=west,font=\bfseries\normalsize,text=figink] at (8.36,7.78) {b};
\node[anchor=west,font=\small\bfseries,text=figink] at (.48,7.28) {One plant, many disconnected descriptions};
\node[anchor=west,font=\small\bfseries,text=figink] at (8.44,7.28) {One change, connected engineering context};
\foreach \x/\y/\title in {.55/5.04/P\&ID,3.00/5.04/Simulation,5.45/5.04/Equipment,.55/1.40/Layout,3.00/1.40/Controls,5.45/1.40/Specifications}{
 \fill[white,rounded corners=3pt] (\x,\y) rectangle ++(1.80,1.50);
 \draw[figmuted!35,line width=.5pt,rounded corners=3pt] (\x,\y) rectangle ++(1.80,1.50);
 \node[font=\fontsize{7}{9}\selectfont\bfseries,text=figink] at (\x+.9,\y+1.21) {\title};
}
\draw[figink,line width=.6pt,rounded corners=2.5pt] (.80,5.27) rectangle (1.04,5.92);
\draw[figink,line width=.6pt] (1.04,5.37)--(2.18,5.37);
\draw[figink,line width=.5pt] (1.27,5.30)--(1.27,5.44)--(1.43,5.30)--(1.43,5.44)--cycle;
\draw[figink,line width=.5pt] (1.35,5.37)--(1.35,5.50);
\draw[figink,line width=.5pt] (1.27,5.50) arc (180:0:.08)--cycle;
\draw[figink,line width=.6pt,fill=white] (1.82,5.37) circle (.12);
\draw[figink,line width=.5pt] (1.82,5.49)--(1.97,5.49);
\draw[figink,line width=.5pt] (1.04,5.80)--(1.24,5.80);
\draw[figink,line width=.5pt,fill=white] (1.35,5.80) circle (.11);
\draw[figink,line width=.4pt] (1.24,5.80)--(1.46,5.80);
\draw[figblue,line width=.5pt,dash pattern=on 1.2pt off 1pt] (1.35,5.69)--(1.35,5.58);
\draw[figmuted,line width=.5pt,-{Latex[length=.9mm]}] (3.24,5.24)--(4.62,5.24);
\draw[figmuted,line width=.5pt,-{Latex[length=.9mm]}] (3.24,5.24)--(3.24,6.02);
\draw[figblue,line width=.9pt] (3.32,5.33)..controls(3.86,5.33) and (3.98,5.86)..(4.50,5.88);
\foreach \px/\py in {3.52/5.40,3.80/5.53,4.02/5.71,4.26/5.80,4.44/5.90} \fill[figink] (\px,\py) circle (.035);
\draw[figink,line width=.7pt] (6.02,5.52) circle (.23);
\draw[figink,line width=.65pt] (5.63,5.52)--(5.79,5.52);
\draw[figink,line width=.65pt] (5.63,5.44)--(5.63,5.60);
\draw[figink,line width=.65pt] (6.12,5.73)--(6.12,5.97) (6.05,5.97)--(6.19,5.97);
\draw[figink,line width=.55pt] (6.25,5.52)--(6.42,5.52);
\draw[figink,line width=.65pt,rounded corners=1pt] (6.42,5.37) rectangle (6.95,5.67);
\draw[figink,line width=.6pt] (5.80,5.23)--(7.02,5.23);
\draw[figink,line width=.5pt] (5.92,5.23)--(5.95,5.30) (6.12,5.23)--(6.09,5.30);
\draw[figmuted,line width=.5pt,dash pattern=on 2pt off 1pt] (.74,1.56) rectangle (2.16,2.38);
\draw[figink,line width=.55pt] (.98,2.13) circle (.12) (.98,1.81) circle (.12);
\draw[figink,line width=.55pt] (1.52,1.98) rectangle (2.04,2.28);
\draw[figblue,line width=.6pt] (1.28,1.70)--(2.04,1.70) (1.28,1.78)--(2.04,1.78);
\draw[figblue,line width=.55pt] (1.10,2.13)--(1.30,2.13)--(1.30,1.78) (1.10,1.81)--(1.20,1.81)--(1.20,1.74)--(1.28,1.74);
\draw[figink,line width=.6pt] (3.20,1.68)--(4.60,1.68);
\draw[figink,line width=.5pt] (3.55,1.68)--(3.55,1.93);
\draw[figink,line width=.5pt,fill=white] (3.55,2.04) circle (.11);
\draw[figink,line width=.4pt] (3.44,2.04)--(3.66,2.04);
\draw[figink,line width=.5pt] (4.12,1.61)--(4.12,1.75)--(4.30,1.61)--(4.30,1.75)--cycle;
\draw[figink,line width=.5pt] (4.21,1.68)--(4.21,1.88);
\draw[figink,line width=.5pt] (4.11,1.88) arc (180:0:.10)--cycle;
\draw[figblue,line width=.5pt,dash pattern=on 1.2pt off 1pt] (3.66,2.04)--(4.21,2.04)--(4.21,1.98);
\fill[figgray] (5.66,2.17) rectangle (7.04,2.34);
\draw[figmuted!80,line width=.45pt] (5.66,1.58) rectangle (7.04,2.34);
\draw[figmuted!80,line width=.35pt] (5.66,2.17)--(7.04,2.17) (5.66,1.97)--(7.04,1.97) (5.66,1.77)--(7.04,1.77) (6.22,1.58)--(6.22,2.17);
\foreach \yy in {2.07,1.87,1.67} \draw[figmuted!70,line width=.6pt] (5.74,\yy)--(6.12,\yy);
\foreach \yy in {2.07,1.67} \draw[figmuted!45,line width=.6pt] (6.30,\yy)--(6.86,\yy);
\draw[figmuted!70,line width=.6pt] (5.74,2.255)--(6.40,2.255);
\fill[figred!14] (6.23,1.78) rectangle (7.03,1.96);
\draw[figred,line width=.6pt] (6.30,1.87)--(6.86,1.87);
\foreach \bx/\by in {4.74/6.49,7.19/6.49}{
 \fill[figred,draw=white,line width=.6pt,rounded corners=.4pt] (\bx,\by+.12)--(\bx-.13,\by-.09)--(\bx+.13,\by-.09)--cycle;
 \node[text=white,font=\fontsize{4.5}{5}\selectfont\bfseries,inner sep=0] at (\bx,\by-.015) {!};
}
\draw[figmuted!60,line width=.9pt] (1.42,3.05)--(6.35,3.05);
\draw[figmuted,line width=.9pt] (2.04,3.05)--(2.04,4.55) arc(180:0:.43 and .13)--(2.90,3.05);
\draw[figmuted,line width=.7pt] (2.04,4.55) arc(180:360:.43 and .13);
\foreach \yy in {3.5,3.9,4.3} \draw[figmuted!65,line width=.35pt] (2.06,\yy)--(2.86,\yy);
\draw[figmuted,line width=1pt] (4.10,3.05)--(4.10,4.27) arc(180:0:.55 and .14)--(5.20,3.05);
\draw[figmuted,line width=.65pt] (4.1,4.27) arc(180:360:.55 and .14);
\draw[figmuted,line width=1pt] (2.90,3.53)--(3.31,3.53)--(3.31,3.31)--(3.85,3.31)--(3.85,3.55)--(4.10,3.55);
\draw[figmuted,line width=.9pt] (5.20,3.78)--(5.90,3.78)--(5.90,3.05);
\draw[figmuted,line width=.7pt] (3.52,3.31) circle (.19);
\draw[figred,line width=1.1pt] (3.42,3.21)--(3.62,3.41) (3.42,3.41)--(3.62,3.21);
\foreach \xa/\ya/\xb/\yb in {1.45/5.04/2.2/4.62,3.92/5.04/3.9/4.58,6.10/5.04/5.18/4.45,1.45/2.90/1.55/3.03,3.92/2.90/3.92/3.03,6.10/2.90/5.95/3.03}{\draw[figred!60,dashed,line width=.75pt] (\xa,\ya)--(\xb,\yb);}
\fill[white,rounded corners=6pt] (8.50,2.63) rectangle (15.48,6.64);
\draw[figblue!20,rounded corners=6pt] (8.50,2.63) rectangle (15.48,6.64);
\node[anchor=west,font=\scriptsize\bfseries,text=figink] at (8.77,6.32) {Versioned engineering model};
\node[anchor=east,font=\fontsize{7}{9}\selectfont,text=figmuted] at (15.12,6.03) {plant revision v1.4};
\node[anchor=west,font=\fontsize{7}{9}\selectfont,text=figmuted] at (8.95,6.03) {Version history};
\draw[figblue!65,line width=1pt] (9.03,5.72)--(14.88,5.72);
\foreach \x in {9.22,10.25,11.27,12.55,13.68,14.70} \fill[figblue] (\x,5.72) circle (.055);
\draw[figink,line width=.85pt] (9.2,4.00)--(9.2,5.13) arc(180:0:.34 and .11)--(9.88,4.00);
\draw[figink,line width=.5pt] (9.2,5.13) arc(180:360:.34 and .11);
\foreach \y in {4.3,4.7} \draw[figblue!40,line width=.55pt] (9.21,\y)--(9.87,\y);
\draw[figink,line width=.85pt] (12.04,4.00)--(12.04,5.05) arc(180:0:.41 and .13)--(12.86,4.00);
\draw[figink,line width=.5pt] (12.04,5.05) arc(180:360:.41 and .13);
\draw[figink,line width=.7pt] (9.88,4.32)--(10.41,4.32)--(10.41,4.08)--(11.07,4.08)--(11.07,4.34)--(12.04,4.34);
\draw[figink,line width=.7pt] (12.86,4.45)--(13.6,4.45)--(13.6,4.00);
\fill[figblue!12] (10.74,4.10) circle (.42);
\draw[figblue,line width=1.3pt] (10.74,4.10) circle (.24);
\draw[figblue,line width=1pt] (10.68,4.02)--(10.88,4.11)--(10.68,4.2)--cycle;
\node[font=\fontsize{7}{9}\selectfont\bfseries,text=figblue] at (10.77,3.56) {P-101};
\draw[figblue!75,line width=.8pt] (10.73,3.28)--(10.73,3.02)--(14.66,3.02);
\foreach \x in {10.75,12.00,13.23,14.52} \fill[figblue] (\x,3.02) circle(.055);
\foreach \x/\t in {8.57/Hydraulics,10.33/Piping,12.09/Controls,13.85/Safety}{
 \fill[white,rounded corners=2.5pt] (\x,1.58) rectangle ++(1.57,.72);
 \draw[figblue!35,rounded corners=2.5pt,line width=.6pt] (\x,1.58) rectangle ++(1.57,.72);
 \fill[figgreen] (\x+.14,1.94) circle(.065);
 \node[font=\fontsize{7}{9}\selectfont,text=figink,anchor=west,inner sep=1pt] at (\x+.22,1.94) {\t};
 \draw[figblue!50,line width=.65pt] (\x+.78,2.30)--(\x+.78,2.63);
}
\draw[figblue,line width=1.35pt,-{Latex[length=3mm]}] (7.58,4.39)--(8.24,4.39);
\end{tikzpicture}
\endgroup}
  \caption{\textbf{From fragmented engineering artefacts to a shared
    Plant-as-Code model.}
    \textbf{a}, Conventional process engineering coordinates P\&IDs,
    simulations, layout, controls and equipment specifications across
    separate tools and documents, where information easily becomes
    outdated or conflicting (red marks); a pump replacement can require
    manual reconciliation. \textbf{b}, A version-controlled semantic
    plant model connects the pump and the affected engineering domains,
    making impact analysis and review traceable. Connected
  representations do not imply that every downstream change is automatic.}
  \label{fig:engineering-as-code}
\end{figure}

\section{Why plant engineering needs a new paradigm}
A typical project spreads its information across requirements
documents, process flow diagrams~(PFDs), piping and instrumentation
diagrams~(P\&IDs), simulation files, 3D layouts, equipment datasheets
and control narratives. Each lives in its own application and file
format. When a pump is replaced, the change has to be carried by hand
into the P\&ID, the hydraulic calculation, the layout, the instrument
list and the control logic~(Fig.~\ref{fig:engineering-as-code}a). In
discrete manufacturing, replacing drawings with a shared digital
model has cut design-to-inspection cycle times by roughly three
quarters~\cite{Hedberg2016DigitalThread}, yet in process engineering
most handovers remain manual.

These workflows were built for human engineers working one step after
another. An AI agent placed into them inherits the same silos. It can
draft a P\&ID quickly, but it cannot tell whether that drawing agrees
with the simulation or the layout, and the engineer reviewing it
cannot either without repeating the cross-checks. The same limit
applies to AI assistants built into individual design tools. A
copilot in a P\&ID editor or a 3D layout package sees only the data
of that tool, so however capable it becomes, its reach ends at the
tool's boundary. A generated document is not a validated design. The
problem is sharper for existing plants, where the current state must
first be reconstructed from scanned drawings and outdated records
before any modification can be assessed~\cite{Theisen2023Digitization}.

\section{Engineering process plants as code}
Software is built from folders of source code and configuration
files. Infrastructure as Code~(IaC) applied the same idea to servers
and networks: computing resources are defined in text files, reviewed
like code and deployed reproducibly~\cite{Morris2016IaC}.
Engineering-as-Code approaches now bring text-based design
representations and automated rule checking to mechanical and
data-centre design~\cite{Song2026EngineeringAsCode}, and open data
platforms bring version control to building
design~\cite{Poinet2020Speckle}. We argue that these principles
should be extended across the lifecycle of process plants, where
designs span many coupled domains and errors have physical
consequences. We refer to this process-specific formulation as
Plant-as-Code~(PaC).

In PaC, the engineering definition of a plant is captured in a
collection of machine-readable files, which together form a
version-controlled representation of the design. Wherever possible,
the files follow existing standards: DEXPI for block flow diagrams,
PFDs and P\&IDs~\cite{Cameron2024DEXPI,Goldstein2025PyDEXPI}, IFC for
the 3D layout and the Module Type Package~(MTP) for interfaces to
package units. The important shift is that these standards define the
data model of a single source of truth, instead of serving as
exchange formats used at the end of a project phase. The files in the
repository need not be the standard exchange files themselves. Large
XML exports produce changes that are hard to review, so a plant
repository may hold the same information in compact text formats such
as YAML that change line by line with the design, and generate the
standard formats whenever data are exchanged. Vendor skids can live
in their own folders with the same structure and be referenced from
the main repository. Standard equipment could be imported the way a
software package manager pulls in libraries, for instance building on
the Asset Administration Shell.

The repository is kept under version control with
Git~(Fig.~\ref{fig:engineering-as-code}b). A team can create a new
branch to explore a different column configuration or a revised
layout without affecting anyone else, and the branch can be validated
on its own. Working on a branch carries little risk, because the
plant of record stays untouched. Merging a branch back into the main
line updates the plant of record and may conflict with work done in
parallel, so every merge passes agentic and human review.

None of this requires process engineers to learn Git. Agents can
handle branching and merging, and graphical interfaces can show both
the files and their changes in readable form, for example as
highlighted differences on familiar P\&ID drawings. What matters is
that the underlying representation is semantic, comparable line by
line and testable, rather than a set of documents that happen to sit
in a repository. A shared repository alone does not make the models
consistent with each other. That also requires stable identifiers,
shared semantics, validated transformations between models and clear
ownership of each part of the design. With appropriate tooling,
provenance, dependencies and change history can then be recorded as
part of normal engineering work.

\begin{figure}[t]
  \centering
  \resizebox{\textwidth}{!}{
\begingroup\renewcommand{\familydefault}{\sfdefault}\normalfont
\begin{tikzpicture}[x=1cm,y=1cm, line cap=round, line join=round, font=\sffamily]
\definecolor{figink}{HTML}{203344}
\definecolor{figmuted}{HTML}{5E7380}
\definecolor{figblue}{HTML}{147E98}
\definecolor{figpale}{HTML}{ECF6F8}
\definecolor{figgray}{HTML}{F3F6F7}
\definecolor{figgreen}{HTML}{24816E}
\definecolor{figred}{HTML}{BA6557}
\definecolor{figgold}{HTML}{C4923C}
\path[use as bounding box] (0,.95) rectangle (16,7.6);
\fill[figgray,rounded corners=5pt] (.3,2.0) rectangle (3.43,7.48);
\fill[figpale,rounded corners=5pt] (3.65,2.0) rectangle (7.52,7.48);
\fill[figgray,rounded corners=5pt] (7.74,2.0) rectangle (11.78,7.48);
\fill[figpale,rounded corners=5pt] (12.00,2.0) rectangle (15.73,7.48);
\node[anchor=west,font=\bfseries\scriptsize,text=figink] at (.58,7.06) {01 / INTENT};
\node[anchor=west,font=\bfseries\scriptsize,text=figink] at (3.95,7.06) {02 / AI PROPOSALS};
\node[anchor=west,font=\bfseries\scriptsize,text=figink] at (8.04,7.06) {03 / VERIFICATION};
\node[anchor=west,font=\bfseries\scriptsize,text=figink] at (12.29,7.06) {04 / APPROVAL};
\fill[white,rounded corners=4pt] (.60,4.37) rectangle (3.10,6.54);
\draw[figmuted!35,rounded corners=4pt] (.60,4.37) rectangle (3.10,6.54);
\node[anchor=west,font=\fontsize{7}{9}\selectfont,text=figmuted] at (.84,6.22) {REQUEST};
\node[anchor=west,font=\bfseries\Large,text=figink] at (.84,5.51) {+20\%};
\node[anchor=west,font=\fontsize{8}{10}\selectfont,text=figink] at (.84,4.98) {capacity target};
\node[anchor=west,font=\fontsize{7}{9}\selectfont,text=figmuted] at (.84,4.64) {Constraints apply};
\fill[figink!85] (1.20,3.67) circle (.17);
\draw[figink,line width=2.3pt] (.97,3.19)..controls(1.03,3.52)and(1.39,3.52)..(1.45,3.19);
\node[anchor=west,font=\fontsize{7}{9}\selectfont,text=figmuted] at (1.66,3.39) {Engineer sets};
\node[anchor=west,font=\fontsize{7}{9}\selectfont,text=figmuted] at (1.66,3.13) {constraints};
\fill[white,rounded corners=4pt] (3.94,4.55) rectangle (7.25,6.54);
\draw[figblue!20,rounded corners=4pt] (3.94,4.55) rectangle (7.25,6.54);
\node[anchor=west,font=\fontsize{8}{10}\selectfont\bfseries,text=figink] at (4.24,6.14) {Candidate revision};
\node[anchor=west,font=\fontsize{7}{9}\selectfont,text=figmuted] at (4.24,5.79) {Provisional design change};
\draw[figink,line width=.8pt] (4.27,5.05)--(4.85,5.05)--(4.85,5.47) arc(180:0:.24 and .07)--(5.33,5.05)--(5.88,5.05);
\draw[figink,line width=.7pt] (4.85,5.47) arc(180:360:.24 and .07);
\draw[figblue,line width=1pt] (5.88,5.05)--(6.57,5.05);
\fill[figblue!18] (6.17,5.05) circle (.28);
\draw[figblue,line width=1.2pt] (6.17,5.05) circle(.19);
\node[font=\fontsize{6}{8}\selectfont,text=figblue] at (6.16,4.74) {P-101};
\node[font=\fontsize{7}{9}\selectfont\bfseries,text=figink] at (5.59,4.13) {Agents propose edits};
\foreach \xx/\txt in {4.20/Process,5.23/Equip.,6.26/Controls}{
 \fill[white,rounded corners=2pt] (\xx,3.23) rectangle ++(.88,.49);
 \draw[figblue!35,rounded corners=2pt] (\xx,3.23) rectangle ++(.88,.49);
 \node[font=\fontsize{6.4}{8}\selectfont,text=figink] at (\xx+.44,3.48) {\txt};
}
\node[font=\fontsize{7}{9}\selectfont,text=figmuted] at (5.60,2.97) {Domain tools and agents};
\node[anchor=west,font=\fontsize{7}{9}\selectfont,text=figmuted] at (8.05,6.45) {No approval without evidence};
\foreach \y/\name/\sub in {5.71/Physics/Flows and constraints,4.77/Engineering/Equipment and controls,3.83/Safety/HAZOP and safeguards}{
 \fill[white,rounded corners=2.5pt] (8.05,\y-.48) rectangle (11.46,\y+.26);
 \draw[figmuted!25,rounded corners=2.5pt] (8.05,\y-.48) rectangle (11.46,\y+.26);
 \draw[figblue,line width=.85pt] (8.28,\y-.10)--(8.39,\y-.20)--(8.58,\y+.08);
 \node[anchor=west,font=\fontsize{8}{9}\selectfont\bfseries,text=figink] at (8.73,\y+.02) {\name};
 \node[anchor=west,font=\fontsize{6.5}{8}\selectfont,text=figmuted] at (8.73,\y-.24) {\sub};
}
\draw[figred,line width=.95pt,-{Latex[length=2mm]}] (8.63,2.86)--(8.63,2.48)--(5.69,2.48)--(5.69,2.98);
\node[fill=figgray,font=\fontsize{7}{9}\selectfont,text=figred] at (6.85,2.38) {Revise on failure};
\fill[white,rounded corners=4pt] (12.29,4.25) rectangle (15.40,6.54);
\draw[figblue!24,rounded corners=4pt] (12.29,4.25) rectangle (15.40,6.54);
\node[font=\fontsize{8}{10}\selectfont\bfseries,text=figink] at (13.84,6.10) {Review evidence};
\node[font=\fontsize{7}{9}\selectfont,text=figmuted] at (13.84,5.73) {Engineer evaluates results};
\draw[figgreen,line width=1.6pt] (13.53,5.15)--(13.74,4.93)--(14.15,5.47);
\node[font=\fontsize{8}{10}\selectfont\bfseries,text=figgreen] at (13.86,4.60) {Engineer authorizes};
\fill[white,rounded corners=3pt] (12.29,2.70) rectangle (15.40,3.90);
\draw[figblue!25,rounded corners=3pt] (12.29,2.70) rectangle (15.40,3.90);
\node[anchor=west,font=\fontsize{8}{10}\selectfont\bfseries,text=figink] at (12.53,3.55) {05 / RELEASE};
\node[anchor=west,font=\fontsize{7}{9}\selectfont,text=figmuted] at (12.53,3.22) {Commit + audit trail};
\draw[figblue,line width=.85pt] (12.55,2.91)--(14.25,2.91);
\foreach \x in {12.63,13.37,14.15} \fill[figblue] (\x,2.91) circle(.07);
\foreach \xa/\xb in {3.40/3.78,7.50/7.88,11.76/12.15}{\draw[figblue,line width=1.1pt,-{Latex[length=2.3mm]}] (\xa,5.45)--(\xb,5.45);}
\draw[figblue,line width=1pt,-{Latex[length=2.0mm]}] (13.87,4.25)--(13.87,3.88);
\draw[figink!45,line width=.85pt,-{Latex[length=2.3mm]}] (13.86,2.53)--(13.86,1.29)--(1.89,1.29)--(1.89,1.87);
\node[fill=white,font=\fontsize{7.4}{9}\selectfont,text=figmuted] at (7.87,1.29) {Operational feedback becomes a new, traceable change request};
\end{tikzpicture}
\endgroup}
  \caption{\textbf{Design-Ops for a hypothetical capacity increase.}
    An engineer requests a 20\% increase in production capacity subject
    to plant constraints. Agents propose coordinated modifications,
    engineering tools independently check physical and equipment
    constraints and safety requirements, and an engineer reviews the
    evidence before authorizing a versioned design change. In the
    example shown, all checks have passed. Failed checks trigger
    revision; operational feedback initiates further traceable
    requests. The workflow is illustrative, not a demonstrated
  autonomous plant deployment.}
  \label{fig:design-ops}
\end{figure}
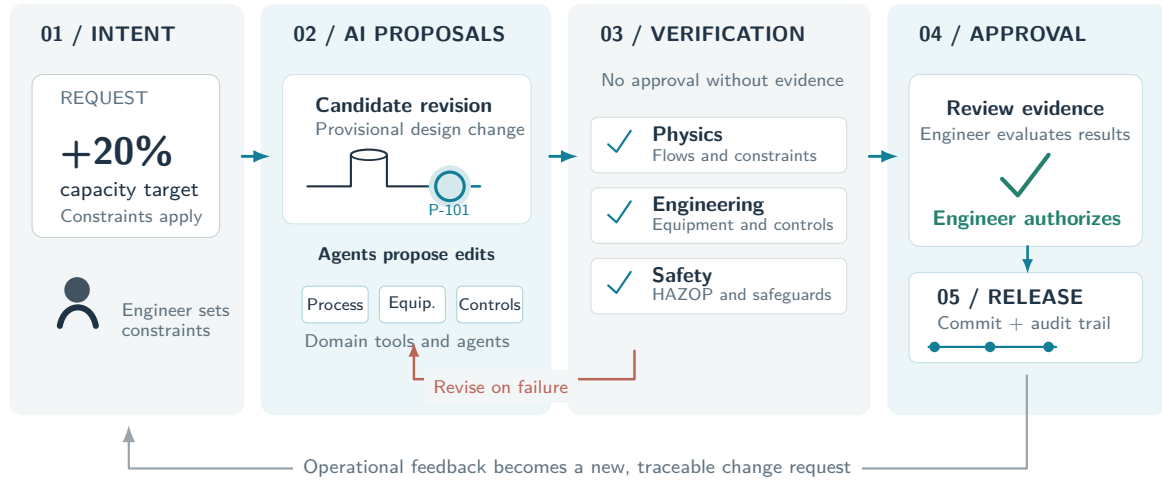

\section{Agentic AI as engineering orchestrator}
On this foundation, agents can support engineers by coordinating much
of the routine work between disciplines, while engineers remain
responsible for design decisions. An agent can break an objective
into tasks, reason about intermediate results and call tools, and
several agents can divide work across
disciplines~\cite{Rupprecht2026MultiAgent}. The calculations
themselves should stay with deterministic, specialized software for
simulation, equipment sizing, pipe routing and control design, as
tool-use architectures have already demonstrated in other domains
such as chemistry~\cite{Bran2024ChemCrow}.

Engineers set the objectives, constraints and design ideas that
agents work on, and they judge the results. They can review these
results in familiar environments rather than in code, for example by
marking up a P\&ID or walking through the plant model in virtual
reality. Engineers can also talk to the agent during the
review~\cite{Alimin2026ChatPID}. They can ask why a line was routed
along a particular path or why a specific instrument was selected,
ask for a pipe rack to be moved, or request a comparison with last
month's version.

The familiar sequence of engineering phases can remain, from concept
through basic to detailed engineering, and from process design to
layout, piping and isometrics. What changes is the duration.
Individual design iterations that take weeks today could take hours.

Current demonstrations cover parts of this chain, such as generating
simulations from text~\cite{Laub2026TextToFlowsheet} or querying
P\&IDs in natural language through knowledge
graphs~\cite{Alimin2026ChatPID}. Beyond academic demonstrations,
early industrial pilots on individual projects, in which one of us
(D.D.) applied version-controlled, agent-assisted Design-Ops
workflows, point to substantially faster design iterations. A
workflow demonstration is still a long way from a construction-ready
design.

\section{From DevOps to Design-Ops}
Many software teams manage the life of their products through DevOps.
Code moves from development into operations, and what is learned in
operation starts the next development cycle. Continuous integration
and delivery make this loop fast and
routine~\cite{Humble2010ContinuousDelivery}. Plant engineering
follows a staged, waterfall-like process for good reasons, since a
plant is designed, procured and built once and its errors are physical.

Building on established management-of-change and digital-thread
practices, Design-Ops extends DevOps principles to engineering
workflows, as previously proposed and piloted for process automation
engineering~\cite{Bacidore2025DesignOps}. Here, we extend this
approach across process plant engineering, introducing
version-controlled changes, automated verification and traceable
dependencies between engineering
disciplines~(Fig.~\ref{fig:design-ops}). Every item in a downstream
step is linked to the upstream information it derives from. A line
size in the piping model points to a flow rate in the heat and mass
balance, which in turn points to a design case. When the design case
changes, the affected items are known immediately.

This rests on a once-only principle, the engineering counterpart of
the ``don't repeat yourself'' rule in software development. A data
point is engineered and defined once and never copied by hand into
another document. Manual re-entry is one of the largest sinks of time
in current projects and a common source of inconsistencies.
Design-Ops also replaces mostly manual checking with automated tests
that run whenever a change is proposed.

Software practice offers proven ways to organize fast and
collaborative work on a shared design, but the analogy has limits. A
faulty software release can often be rolled back within minutes,
whereas a wrongly specified vessel may already be fabricated. We
therefore treat software practice as a template for workflow and not
as evidence of safety. Design-Ops must embed management of change,
approval by accountable engineers and feedback from operations, so
that what operators learn starts a new and traceable change request.

\section{Verification is the limiting step}
Agents can propose designs faster than anyone can check them.
Reliable agentic engineering therefore depends on a closed loop in
which verification results flow back to the agent. When an automated
check reports that a pipe crosses a walkway, the agent reroutes it
and the check runs again.

Verification has several layers. Syntactic checks confirm that files
conform to their schema. Semantic checks confirm, for example, that a
valve on the P\&ID also exists in the instrument list and in the 3D
model. Physical checks cover mass and energy balances, hydraulics and
equipment limits, and safety checks cover relief cases, hazardous
areas and separation distances. Every proposed change should state
its preconditions, constraints and expected outcomes, and report
uncertainty, so that reviewers can see which checks ran and which did
not. Effects across domains deserve particular attention. For
example, rerouting a line changes its pressure drop, which can change
the pump duty and the sizing of a control valve.

Some qualities are hard to formalize. Experienced engineers recognize
a neat, maintainable piping layout at a glance, yet it is difficult
to write down as a rule. Reviewer agents can help with such gaps, but
they are slower and less predictable than deterministic checks, and
an agent built on the same model as the designing agent is not truly
independent. Their output is evidence for a human reviewer, not
verification in itself. Human review is slower and more expensive
still, but it remains indispensable. Experienced engineers bring
judgement and accountability that automated checks cannot provide.
Regulatory frameworks such as the EU AI Act also require human
oversight of AI systems classified as high-risk. We therefore expect
a bounded autonomy in which agents propose, tools check, engineers
review and authorize, and only then is a change released.
Agent-computer interfaces from software engineering show how
automated checks can be built into an agent's working
loop~\cite{Yang2024SWEAgent}, although they tell us nothing yet about
plant safety.

The field also needs benchmarks. SWE-bench made the progress of
software agents measurable~\cite{Jimenez2024SWEBench}. Plant
engineering needs shared test cases that report engineering errors,
the number of human interventions and the effort required to repair a design.

Over time, agents may take on more of the verification work, but only
where evidence shows that they perform at least as well as the checks
they replace and within limits that regulators and operators accept.
As that evidence accumulates, some modifications could be engineered
in days rather than months.

\section{Outlook}
If designs can be produced and checked much faster, exploring
alternatives, running value engineering studies and optimizing the
plant become cheap. The bottleneck then moves to construction, and
engineering effort will shift towards making plants faster and
cheaper to build, for example through modular units fabricated off site.

The consequences reach beyond engineering tools. Engineering
contractors, whose revenue often depends on engineering hours, may
increasingly sell verified designs and controlled changes instead of
time. When engineering knowledge is captured in shared models, tools
and checks, it becomes easier to transfer, giving regions with few
experienced engineers access to expertise that is scarce today.
Faster engineering could also shorten the time needed to deploy the
new processes that the energy and materials transitions require, or
to scale up the production of medicines in a crisis such as the
COVID-19 pandemic.

These changes also carry risks. Junior engineers learn much of their
craft through routine checking and data transfer, the very work that
agents may take over. Without new ways to build that experience, the
profession could lose the judgement it needs. Engineers may also come
to trust agent output too readily, and an error in a shared template
or check could spread to many designs at once. Machine-readable plant
models are attractive targets for cyberattacks and raise new
questions of intellectual property. And if a few software providers
control the models and agents, dependence on them could replace
today's fragmentation and create single points of failure that reduce
the robustness of the whole industry.

Turning Plant-as-Code into engineering practice requires effort from
academia, industry and policymakers alike. Although much of it
resembles software development, Plant-as-Code raises fundamental
research questions for academia. How can engineering semantics be
shared across disciplines so that a pump means the same thing to the
process, piping and automation engineer? How can agent-proposed
changes be verified with quantified confidence, and which open
benchmarks reflect real engineering work on new and existing
plants~\cite{Rupprecht2026MultiAgent,Jimenez2024SWEBench}? And how
should engineers be trained to direct and supervise agents? Industry,
in turn, needs open tool interfaces for agents, and policymakers and
regulatory bodies must set clear rules on liability, accountability
and the certification of AI-assisted designs.

The move towards Engineering-as-Code is a chance to rethink how
process plants are designed, modified and operated. For process
engineering, the central challenge is to connect heterogeneous
engineering models into one coherent and verifiable description of
the plant. With Plant-as-Code as this foundation and Design-Ops
organizing controlled changes, agents could increasingly coordinate
design work while engineers keep responsibility for consequential decisions.

\section*{Acknowledgements}
This research is supported by the Netherlands Organisation for Scientific Research (NWO) under the
Veni grant (ID 20196), for which we express sincere gratitude.

\section*{Declaration of AI use}
The authors used GPT-6 (OpenAI) and Claude Opus 5.5 (Anthropic) to
assist with drafting, literature searches, language editing and
figure generation. The authors independently developed the scientific
concepts, arguments and conclusions, critically evaluated and revised
all AI-assisted content, verified the cited sources and retained full
intellectual and editorial control over the manuscript. They take
full responsibility for its final content.

\section*{Competing interests}
Davy Demeyer is the founder of Acceleer, a company developing
commercial software for AI-assisted plant engineering.

\bibliographystyle{unsrtnat}
\bibliography{references,iac-references}

\begin{thebibliography}{15}
\providecommand{\natexlab}[1]{#1}
\providecommand{\url}[1]{\texttt{#1}}
\expandafter\ifx\csname urlstyle\endcsname\relax
  \providecommand{\doi}[1]{doi: #1}\else
  \providecommand{\doi}{doi: \begingroup \urlstyle{rm}\Url}\fi

\bibitem[Laub et~al.(2026)Laub, Bosetti, and Bardow]{Laub2026TextToFlowsheet}
Jan-Frederic Laub, Luca Bosetti, and Andr{\'e} Bardow.
\newblock Text-to-flowsheet: an {LLM}-assisted pipeline for expert-level
  digitization and automated simulation of chemical processes.
\newblock \emph{Digital Discovery}, 5\penalty0 (7):\penalty0 2937--2951, 2026.
\newblock \doi{10.1039/D6DD00060F}.

\bibitem[Rupprecht et~al.(2026)Rupprecht, Gao, Karia, and
  Schweidtmann]{Rupprecht2026MultiAgent}
Sophia Rupprecht, Qinghe Gao, Tanuj Karia, and Artur~M. Schweidtmann.
\newblock Multi-agent systems for chemical engineering: a review and
  perspective.
\newblock \emph{Current Opinion in Chemical Engineering}, 51:\penalty0 101209,
  2026.
\newblock \doi{10.1016/j.coche.2025.101209}.

\bibitem[Hedberg et~al.(2016)Hedberg, Lubell, Fischer, Maggiano, and
  Feeney]{Hedberg2016DigitalThread}
Thomas~D. Hedberg, Joshua Lubell, Lyle Fischer, Larry Maggiano, and
  Allison~Barnard Feeney.
\newblock Testing the digital thread in support of model-based manufacturing
  and inspection.
\newblock \emph{Journal of Computing and Information Science in Engineering},
  16\penalty0 (2):\penalty0 021001, 2016.
\newblock \doi{10.1115/1.4032697}.

\bibitem[Theisen et~al.(2023)Theisen, Flores, Schulze~Balhorn, and
  Schweidtmann]{Theisen2023Digitization}
Maximilian~F. Theisen, Kenji~Nishizaki Flores, Lukas Schulze~Balhorn, and
  Artur~M. Schweidtmann.
\newblock Digitization of chemical process flow diagrams using deep
  convolutional neural networks.
\newblock \emph{Digital Chemical Engineering}, 6:\penalty0 100072, 2023.
\newblock \doi{10.1016/j.dche.2022.100072}.

\bibitem[Morris(2016)]{Morris2016IaC}
Kief Morris.
\newblock \emph{Infrastructure as Code: Managing Servers in the Cloud}.
\newblock O'Reilly Media, 2016.
\newblock ISBN 9781491924358.

\bibitem[Song(2026)]{Song2026EngineeringAsCode}
Difei Song.
\newblock Engineering as code: Bringing software engineering methodology to
  engineering design.
\newblock Preprint at \url{https://arxiv.org/abs/2609.04216}, 2026.

\bibitem[Poinet et~al.(2020)Poinet, Stefanescu, and
  Papadonikolaki]{Poinet2020Speckle}
Paul Poinet, Dimitrie Stefanescu, and Eleni Papadonikolaki.
\newblock Collaborative workflows and version control through open-source and
  distributed common data environment.
\newblock In \emph{Proceedings of the 18th International Conference on
  Computing in Civil and Building Engineering}, volume~98 of \emph{Lecture
  Notes in Civil Engineering}, pages 228--247. Springer, 2020.
\newblock \doi{10.1007/978-3-030-51295-8_18}.

\bibitem[Cameron et~al.(2024)Cameron, Otten, Temmen, Hole, and
  Tolksdorf]{Cameron2024DEXPI}
David~B. Cameron, Wilhelm Otten, Heiner Temmen, Monica Hole, and Gregor
  Tolksdorf.
\newblock {DEXPI} process: Standardizing interoperable information for process
  design and analysis.
\newblock \emph{Computers \& Chemical Engineering}, 182:\penalty0 108564, 2024.
\newblock \doi{10.1016/j.compchemeng.2023.108564}.

\bibitem[Goldstein et~al.(2025)Goldstein, Schulze~Balhorn, Alimin, and
  Schweidtmann]{Goldstein2025PyDEXPI}
Dominik~P. Goldstein, Lukas Schulze~Balhorn, Achmad~Anggawirya Alimin, and
  Artur~M. Schweidtmann.
\newblock {pyDEXPI}: A {Python} framework for piping and instrumentation
  diagrams ({P\&IDs}) using the {DEXPI} information model.
\newblock \emph{Systems and Control Transactions}, 4:\penalty0 1365--1370,
  2025.
\newblock \doi{10.69997/sct.139043}.

\bibitem[Bran et~al.(2024)Bran, Cox, Schilter, Baldassari, White, and
  Schwaller]{Bran2024ChemCrow}
Andres~M. Bran, Sam Cox, Oliver Schilter, Carlo Baldassari, Andrew~D. White,
  and Philippe Schwaller.
\newblock Augmenting large language models with chemistry tools.
\newblock \emph{Nature Machine Intelligence}, 6:\penalty0 525--535, 2024.
\newblock \doi{10.1038/s42256-024-00832-8}.

\bibitem[Alimin and Schweidtmann(2026)]{Alimin2026ChatPID}
Achmad~Anggawirya Alimin and Artur~M. Schweidtmann.
\newblock {GraphRAG} for engineering diagrams: {ChatP\&ID} enables {LLM}
  interaction with {P\&IDs}.
\newblock \emph{AIChE Journal}, 72\penalty0 (10):\penalty0 e70540, 2026.
\newblock \doi{10.1002/aic.70540}.

\bibitem[Humble and Farley(2010)]{Humble2010ContinuousDelivery}
Jez Humble and David Farley.
\newblock \emph{Continuous Delivery: Reliable Software Releases through Build,
  Test, and Deployment Automation}.
\newblock Addison-Wesley Professional, 2010.
\newblock ISBN 9780321601919.

\bibitem[Bacidore(2025)]{Bacidore2025DesignOps}
Mike Bacidore.
\newblock Agentic {AI} needs {SDA} to bring a faster pace and more innovation
  to the world of automation.
\newblock Control Design,
  \url{https://www.controldesign.com/design/development-platforms/article/55270321/agentic-ai-needs-sda-to-bring-a-faster-pace-and-more-innovation-to-the-world-of-automation},
  2025.
\newblock Interview with D. Demeyer, 24 February 2025.

\bibitem[Yang et~al.(2024)Yang, Jimenez, Wettig, Lieret, Yao, Narasimhan, and
  Press]{Yang2024SWEAgent}
John Yang, Carlos~E. Jimenez, Alexander Wettig, Kilian Lieret, Shunyu Yao,
  Karthik~R. Narasimhan, and Ofir Press.
\newblock {SWE}-agent: Agent-computer interfaces enable automated software
  engineering.
\newblock In \emph{Advances in Neural Information Processing Systems},
  volume~37, pages 50528--50652, 2024.
\newblock \doi{10.52202/079017-1601}.

\bibitem[Jimenez et~al.(2024)Jimenez, Yang, Wettig, Yao, Pei, Press, and
  Narasimhan]{Jimenez2024SWEBench}
Carlos~E. Jimenez, John Yang, Alexander Wettig, Shunyu Yao, Kexin Pei, Ofir
  Press, and Karthik~R. Narasimhan.
\newblock {SWE}-bench: Can language models resolve real-world {GitHub} issues?
\newblock In \emph{International Conference on Learning Representations}, 2024.

\end{thebibliography}
\end{document}